\documentclass[12pt,english,draftcls, one column]{IEEEtran}
\usepackage[T1]{fontenc}
\usepackage[latin9]{inputenc}
\usepackage{float}
\usepackage{amsthm}
\usepackage{amsmath}
\usepackage{amssymb}
\usepackage{graphicx}

\makeatletter

\floatstyle{ruled}
\newfloat{algorithm}{tbp}{loa}
\providecommand{\algorithmname}{Algorithm}
\floatname{algorithm}{\protect\algorithmname}

\theoremstyle{plain}

\theoremstyle{plain}

\ifCLASSINFOpdf
\else
\fi
\usepackage{babel}

\makeatother

\usepackage{babel}
\providecommand{\propositionname}{Proposition}
\providecommand{\theoremname}{Theorem}

\begin{document}

\title{Multi-Layer Transmission and Hybrid Relaying for Relay Channels with\\
Multiple Out-of-Band Relays}

\author{Seok-Hwan Park, Osvaldo Simeone, Onur Sahin and Shlomo Shamai (Shitz)
\thanks{S.-H. Park and O. Simeone are with the Center for Wireless Communications
and Signal Processing Research (CWCSPR), ECE Department, New Jersey
Institute of Technology (NJIT), Newark, NJ 07102, USA (email: \{seok-hwan.park,
osvaldo.simeone\}@njit.edu).

O. Sahin is with InterDigital Inc., Melville, New York, 11747, USA
(email: Onur.Sahin@interdigital.com).

S. Shamai (Shitz) is with the Department of Electrical Engineering,
Technion, Haifa, 32000, Israel (email: sshlomo@ee.technion.ac.il).%
}}
\maketitle
\begin{abstract}
In this work, a relay channel is studied in which a source encoder
communicates with a destination decoder through a number of out-of-band
relays that are connected to the decoder through capacity-constrained
digital backhaul links. This model is motivated by the uplink of cloud
radio access networks. In this scenario, a novel transmission and
relaying strategies are proposed in which multi-layer transmission
is used, on the one hand, to adaptively leverage the different decoding
capabilities of the relays and, on the other hand, to enable hybrid
decode-and-forward (DF) and compress-and-forward (CF) relaying. The
hybrid relaying strategy allows each relay to forward part of the
decoded messages and a compressed version of the received signal to
the decoder. The problem of optimizing the power allocation across
the layers and the compression test channels is formulated. Albeit
non-convex, the derived problem is found to belong to the class of
so called complementary geometric programs (CGPs). Using this observation,
an iterative algorithm based on the homotopy method is proposed that
achieves a stationary point of the original problem by solving a sequence
of geometric programming (GP), and thus convex, problems. Numerical
results are provided that show the effectiveness of the proposed multi-layer
hybrid scheme in achieving performance close to a theoretical (cutset)
upper bound.

~~

~~\end{abstract}
\theoremstyle{theorem}
\newtheorem{theorem}{Theorem}
\theoremstyle{proposition}
\newtheorem{proposition}{Proposition}
\theoremstyle{lemma}
\newtheorem{lemma}{Lemma}
\theoremstyle{corollary}
\newtheorem{corollary}{Corollary}
\theoremstyle{definition}
\newtheorem{definition}{Definition}
\theoremstyle{remark}
\newtheorem{remark}{Remark}

\section{Introduction}

\begin{figure}
\centering\includegraphics[width=12cm,height=6cm]{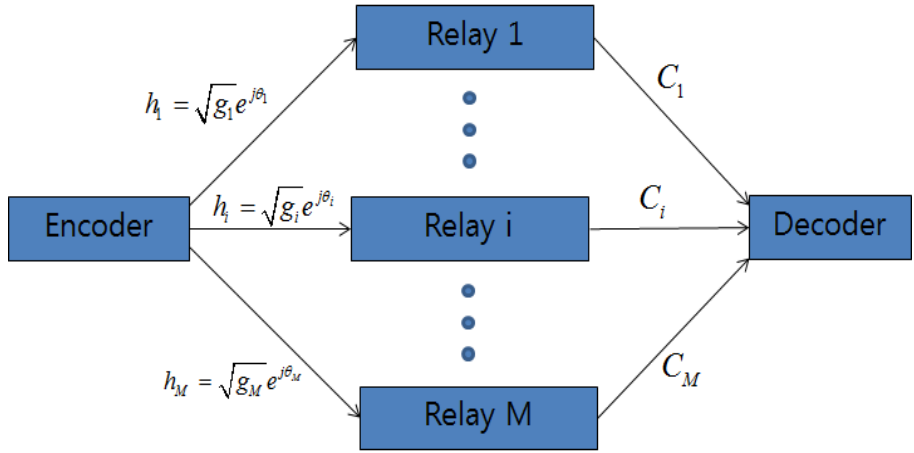}

\caption{\label{fig:block diagram}Illustration of the considered channel with
multiple relays connected to the decoder via out-of-band digital backhaul
links with given capacities.}
\end{figure}

The multiple relay network, in which a source encoder wishes to communicate
with a destination through a number of relays, as seen in Fig. \ref{fig:block diagram},
has been actively studied due to its wide range of applications. Most
of the activity, starting from \cite{Schein}, focuses on Gaussian
networks in which the first hop amounts to a Gaussian broadcast channel
from source to relays and the second hop to a multiple access channel
between relays and receivers. The literature on this subject is vast
and includes the proposal of various transmission strategies, including
\textit{decode-and-forward} (DF) \cite{Schein}-\cite{dCoso relay},
\textit{compress-and-forward} (CF) \cite{Schein}-\cite{WeiYu}, \textit{amplify-and-forward}
(AF) \cite{Xue}\cite{dCoso relay}\cite{Khandani} and hybrid AF-DF
\cite{Xue}\cite{Khandani}.

In this paper, we are concerned with a variation of the more classical
multi-relay channel discussed above in which the relays are connected
to the destination through digital backhaul links of finite-capacity.
The motivation for this model comes from the application to so called
cloud radio cellular networks, in which the base stations (BSs) act
as relays connected to the central decoder via finite-capacity backhaul
links \cite{Marsch}\cite{Alcatel}. This model was studied in \cite{Sanderovich:08}-\cite{Park}\cite{WeiYu}\cite{Nazer}
(see also review in \cite{Simeone}). References \cite{Sanderovich:08}\cite{dCoso}\cite{Park}\cite{WeiYu}
focus on CF strategies, while \cite{Sanderovich:09} considers hybrid
DF-CF strategies and \cite{Nazer} studies schemes based on \textit{compute-and-forward}.

\subsection{Contributions}

In this paper, we propose a novel transmission and relaying strategy
in which multi-layer transmission is used, on the one hand, in order
to properly leverage the different decoding capabilities of the relays
similar to \cite{Xue}, and, on the other hand, to enable hybrid DF
and CF relaying. In the proposed hybrid relaying strategy, each relay
forwards part of the decoded messages and a compressed version of
the received signal. The multi-layer strategy is designed so as to
facilitate decoding at the destination based on the information received
from the relays. To this end, the proposed design is different from
the classical broadcast coding approach of \cite{Shamai BC} in which
each layer encodes an independent message. Instead, in the proposed
scheme, each layer encodes an appropriately selected set of independent
messages. It is emphasized that the hybrid DF-CF approach studied
in \cite{Sanderovich:09} is based on single-layer transmission.

The problem of optimizing the power allocation across the layers and
the compression test channels is formulated. Albeit non-convex, the
derived problem is found to belong to the class of so called complementary
geometric programs (CGPs) (see \cite[Sec. 3.2]{Codreanu} for more
detail). Using this observation, an iterative algorithm based on the
homotopy method is proposed that achieves a stationary point of the
original problem by solving a sequence of geometric programming (GP)
\cite{Boyd}, and thus convex, problems. Numerical results are provided
that show the effectiveness of the proposed multi-layer hybrid scheme
in achieving performance close to a theoretical cutset upper bound
\cite[Theorem 1]{Cover}.

\textit{Notation}: We use $p(y|x)$ to denote conditional probability
density function (pdf) of random variable $X$ given $Y$. All logarithms
are in base two unless specified. Given a sequence $X_{1},\ldots,X_{m}$,
we define a set $X_{\mathcal{S}}=\{X_{j}|j\in\mathcal{S}\}$ for a
subset $\mathcal{S}\subseteq\{1,\ldots,m\}$; we set $X_{\phi}$ as
the empty set.

\section{System Model\label{sec:System-Model}}

We consider a relay channel in which a source encoder wishes to communicate
with a destination decoder through a number $M$ of relays as illustrated
in Fig. \ref{fig:block diagram}. We denote the set of relays by $\mathcal{M}=\{1,\ldots,M\}$.
The relays operate out of band in the sense that each $i$th relay
is connected to the receiver via an orthogonal finite-capacity link
of capacity $C_{i}$ in bits per channel use (c.u.). The encoder transmits
a signal $X$ which is subject to power constraint $\mathbb{E}[|X|^{2}]\leq P$.
Each relay $i$ receives a signal $Y_{i}$ which is given as
\begin{equation}
Y_{i}=h_{i}X+Z_{i}\label{eq:received signals}
\end{equation}
with a complex channel coefficient $h_{i}=\sqrt{g_{i}}e^{j\theta_{i}}$
and independent additive white Gaussian noise (AWGN) $Z_{i}\sim\mathcal{CN}(0,1)$
for $i=1,\ldots,M$. We assume that the channel coefficients $h_{1},\ldots,h_{M}$
are constant over a transmission block and are perfectly known to
all nodes. Without loss of generality, the channel powers $g_{1},\ldots,g_{M}$
are assumed to be sorted such that
\begin{equation}
g_{1}\leq\ldots\leq g_{M}.\label{eq:sorted channels}
\end{equation}

\section{Multi-Layer Transmission with Hybrid Relaying\label{sec:Proposed}}

In this section, we propose a transmission strategy that is based
on multi-layer transmission and hybrid relaying. Hybrid relaying is
performed by having each relay forward part of the decoded messages,
which amounts to partial decode-and-forward (DF), along with a compressed
version of the received signal, thus adhering also to the compress-and-forward
(CF) paradigm. The multi-layer strategy used at the source is designed
so as to facilitate decoding at the destination based on the information
received from the relays, as detailed below.

\subsection{Multi-Layer Transmission\label{sub:Broadcast coding}}

The amount of information decodable at the relays depends on the generally
different fading powers $g_{1},\ldots,g_{M}$. To leverage the different
channel qualities, we enable flexible decoding at the relays by adopting
a multi-layer transmission strategy at the encoder. This approach
was also considered in \cite{Xue} for the case of two relays that
communicate to the decoder via multiple access Gaussian channels.
We assume that the transmitter splits its message into $M+1$ independent
submessages, say $W_{1},\ldots,W_{M+1}$, with corresponding rates
$R_{1},\ldots,R_{M+1}$ in bit/c.u., respectively. The idea is that
message $W_{1}$ will be decoded by all relays, message $W_{2}$ only
by relays $2,\ldots,M$, and so on. This way, relays with better channel
conditions decode more information. Message $W_{M+1}$ is instead
decoded only at the destination.

To encode these messages, the encoded signal is given by
\begin{equation}
X=\sum_{k=1}^{M+1}\sqrt{P_{k}}X_{k},\label{eq:broadcast coding}
\end{equation}
where the signals $X_{1},\ldots,X_{M+1}$ are independent and distributed
as $\mathcal{CN}(0,1)$, and the power coefficients $P_{1},\ldots,P_{M+1}$
are subject to the power constraint $\sum_{k=1}^{M+1}P_{k}\leq P$.
The signal $X_{1}$ encodes message $W_{1}$, signal $X_{2}$ encodes
both message $W_{1}$ and $W_{2}$, and so on, so that signal $X_{k}$
encodes messages $W_{1},\ldots,W_{k}$ for $k=1,\ldots,M$. Note that,
unlike classical multi-layer transmission \cite{Shamai BC}\cite{Cover BC},
here signal $X_{k}$ does not only encode message $W_{k}$. The reason
for this choice will be clarified below. Finally, signal $X_{M+1}$
encodes message $W_{M+1}$.

Relay 1 decodes message $W_{1}$ from $X_{1}$; relay 2 first decodes
message $W_{1}$ from $X_{1}$ and then message $W_{2}$ from $X_{2}$
using its knowledge of $W_{1}$; and so on, so that relay $k$ decodes
messages $W_{1},\ldots,W_{k}$ for $k=1,\ldots,M$. From standard
information-theoretic considerations, the following conditions are
sufficient to guarantee that rates $R_{k}$ are decodable by the relays
\cite{Shamai BC}
\begin{equation}
R_{k}\leq I\left(X_{k};Y_{k}|X_{1},\ldots,X_{k-1}\right),\label{eq:DF rate}
\end{equation}
for $k=1,\ldots,M$. This is because, by (\ref{eq:broadcast coding}),
condition (\ref{eq:DF rate}) with $k=1$, namely $R_{1}\leq I(X_{1};Y_{1})$
ensures that not only relay 1 but all relays can decode message $W_{1}$;
and, generalizing, the inequality (\ref{eq:DF rate}) for a given
$k$ guarantees that not only relay $k$ can decode message $W_{k}$
after having decoded $W_{1},\ldots,W_{k-1}$, but also all relays
$k+1,\ldots,M$ can. The signal $X_{M+1}$, and thus message $W_{M+1}$
is decoded by the destination only as it will be described in the
next subsection.

\subsection{Hybrid Relaying\label{sub:Hybrid relaying}}

As discussed, relay $i$ decodes messages $W_{1},\ldots,W_{i}$. Then,
each $i$th relay transmits \textit{partial} information about the
decoded messages to the destination via the backhaul links. The rate
at which this partial information is transmitted to the destination
is selected so as to enable the latter to decode messages $W_{1},\ldots,W_{M}$
jointly based on all the signals received from the relays. This step
will be detailed below. We denote as $C_{i}^{\mathrm{DF}}\leq C_{i}$
the portion of the backhaul capacity devoted to the transmission of
the messages decoded by relay $i$.

Beside the rate allocated to the transmission of (part of) the decoded
messages, relay $i$ utilizes the residual backhaul link to send a
compressed version $\hat{Y}_{i}$ of the received signal $Y_{i}$.
The compression strategy at relay $i$ is characterized by the test
channel $p(\hat{y}_{i}|y_{i})$ according to conventional rate-distortion
theory arguments (see, e.g., \cite{ElGamal}). Moreover, since the
received signals at different relays are correlated with each other,
it is beneficial to adopt a distributed source coding strategy. Here,
similar to \cite{Park}\cite{WeiYu}\cite{Zhang}, we use successive
decoding via Wyner-Ziv compression with a given order $\hat{Y}_{\pi(1)}\rightarrow\ldots\rightarrow\hat{Y}_{\pi(M)}$,
where $\pi(i)$ is a given permutation of the relays' indices $\mathcal{M}$.
Thus, the decoder can successfully retrieve the descriptions $\hat{Y}_{1},\ldots,\hat{Y}_{M}$
if the conditions \cite{WynerZiv}
\begin{equation}
I\left(Y_{\pi(i)};\hat{Y}_{\pi(i)}|\hat{Y}_{\{\pi(1),\ldots,\pi(i-1)\}}\right)\leq C_{\pi(i)}^{\mathrm{CF}}\label{eq:backhaul constraint CF}
\end{equation}
are satisfied for all $i=1,\ldots,M$, where we defined $C_{i}^{\mathrm{CF}}\leq C_{i}$
as the capacity allocated by relay $i$ to communicate the compressed
received signal $\hat{Y}_{i}$ to the decoder. It is recalled that
(\ref{eq:backhaul constraint CF}) is the rate needed to compress
$Y_{\pi(i)}$ as $\hat{Y}_{\pi(i)}$ given that the destination has
side information given by the previously decompressed signals $\hat{Y}_{\pi(1)},\ldots,\hat{Y}_{\pi(i-1)}$.

Without claim of optimality, we assume Gaussian test channel $p(\hat{y}_{i}|y_{i})$,
so that the compressed signal $\hat{Y}_{i}$ can be expressed as
\begin{equation}
\hat{Y}_{i}=Y_{i}+Q_{i},\label{eq:Gaussian test channel}
\end{equation}
where the compression noise $Q_{i}\sim\mathcal{CN}(0,\sigma_{i}^{2})$
is independent of the received signal $Y_{i}$ to be compressed. We
observe that assumption of the Gaussian test channels (\ref{eq:Gaussian test channel})
does not involve any loss of optimality if the relays are allowed
to perform only the CF strategy \cite{dCoso}\cite{bottleneck}\cite{Tian}.
We remark that the compression strategy (\ref{eq:Gaussian test channel})
at relay $i$ is characterized by a single parameter $\sigma_{i}^{2}$.

\subsection{Decoding\label{sub:Decoding-Operation}}

The destination decoder is assumed to first recover the descriptions
$\hat{Y}_{1},\ldots,\hat{Y}_{M}$ from the signals received by the
relays. This step is successful as long as conditions (\ref{eq:backhaul constraint CF})
are satisfied. Having obtained $\hat{Y}_{\mathcal{M}}=\{\hat{Y}_{1},\ldots,\hat{Y}_{M}\}$,
the destination decodes jointly the messages $W_{1},\ldots,W_{M}$
based on the partial information about these messages received from
the relays and on the compressed received signals $\hat{Y}_{\mathcal{M}}$.
Finally, message $W_{M+1}$ is decoded. The following lemma describes
the set of tuples $(R_{1},\ldots,R_{M+1})$ that is achievable via
this strategy.

\begin{lemma}\label{lem:achievability}

A rate tuple $(R_{1},\ldots,R_{M+1})$ is achievable by the proposed
multi-layer strategy with hybrid relaying if the following conditions
are satisfied for some values of $C_{i}^{\mathrm{DF}}\in[0,C_{i}]$,
$i=1,\ldots,M$:
\begin{align}
R_{i} & \leq I\left(X_{i};Y_{i}|X_{1},\ldots,X_{i-1}\right),\,\, i=1,\ldots,M,\label{eq:achievable DF hop1}\\
C_{\pi(i)}^{\mathrm{DF}} & +I\left(Y_{\pi(i)};\hat{Y}_{\pi(i)}|\hat{Y}_{\{\pi(1),\ldots,\pi(i-1)\}}\right)\leq C_{\pi(i)},\,\, i=1,\ldots,M,\label{eq:backhaul CF}\\
\sum_{j=k}^{M}R_{j} & \leq\sum_{j=k}^{M}C_{j}^{\mathrm{DF}}+I\left(X_{\{k,\ldots,M\}};\hat{Y}_{\mathcal{M}}|X_{\{1,\ldots,k-1\}}\right),\,\, k=1,\ldots,M,\label{eq:backhaul DF}\\
\mathrm{and}\,\,\,\, R_{M+1} & \leq I\left(X_{M+1};\hat{Y}_{\mathcal{M}}|X_{\mathcal{M}}\right).\label{eq:achievable CF}
\end{align}

\begin{proof} \emph{The constraint (\ref{eq:achievable DF hop1})
corresponds to (\ref{eq:DF rate}) and guarantees correct decoding
at the relays. Constraint (\ref{eq:backhaul CF}) follows from (\ref{eq:backhaul constraint CF})
and the backhaul constraint. The inequalities in (\ref{eq:backhaul DF})
ensure that the messages $W_{1},\ldots,W_{M}$ are correctly decoded
by the destination based on the partial information received from
the relays and the compressed signals $\hat{Y}_{\mathcal{M}}$. This
is a consequence of well-known results on the capacity of multiple
access channels with transmitters encoding given subsets of messages
\cite{Bruyn} (see also \cite{Gunduz}), as recalled in Appendix \ref{appendix:proof achievability}.
We observe here that the sufficiency of (\ref{eq:backhaul DF}) for
correct decoding hinges on the fact that signal $X_{k}$ encodes messages
$W_{1},\ldots,W_{k}$ for $k=1,\ldots,M$, and not merely $W_{k}$
as in the more conventional multi-layer approach \cite{Cover BC}\cite{Shamai BC}.
Finally, constraint (\ref{eq:achievable CF}) ensures the correct
decoding of message $W_{M+1}$ based on all the decoded signals $X_{\mathcal{M}}$
and the compressed received signals $\hat{Y}_{\mathcal{M}}$.}

\end{proof}

\end{lemma}

\section{Optimization\label{sec:optimization}}

In this section, we are interested in optimizing the power allocation
$P_{1},\ldots,P_{M+1}$, the compression test channels characterized
by the compression noise variances $\sigma_{1}^{2},\ldots,\sigma_{M}^{2}$
and the backhaul capacity allocation between DF and CF relaying, with
the aim of maximizing the sum-rate $R_{\mathrm{sum}}=\sum_{k=1}^{M+1}R_{k}$.
Based on Lemma \ref{lem:achievability}, this problem is formulated
as
\begin{align}
\underset{\begin{array}{c}
\pi,\{P_{k},R_{k}\geq0\}_{k=1}^{M+1},\\
\{\sigma_{i}^{2},C_{i}^{\mathrm{DF}}\geq0\}_{i=1}^{M}
\end{array}}{\mathrm{maximize}} & \sum_{k=1}^{M+1}R_{k}\label{eq:problem original}\\
\mathrm{s.t.}\,\,\,\,\, & \mathrm{(}\ref{eq:achievable DF hop1}\mathrm{)-(}\ref{eq:achievable CF}\mathrm{)},\nonumber \\
 & \sum_{k=1}^{M+1}P_{k}\leq P.\nonumber
\end{align}
In (\ref{eq:problem original}), the optimization space includes the
ordering $\pi$ used for decompression at the decoder, along with
the mentioned power and backhaul allocations and the compression noises.
Due to the inclusion of the ordering $\pi$, the problem is combinatorial.
Therefore, in this section, we focus on the optimization of the other
variables for fixed ordering $\pi$. Optimization of $\pi$ will then
have to be generally performed using an exhaustive search procedure
or using a suitable heuristic method.

Under the assumption of the multi-layer transmission (\ref{eq:broadcast coding}),
the Gaussian test channels (\ref{eq:Gaussian test channel}) and given
ordering $\pi$, the problem (\ref{eq:problem original}) can be written
as\begin{subequations}\label{eq:problem computed}
\begin{align}
\underset{\begin{array}{c}
\{R_{i},C_{i}^{\mathrm{DF}}\geq0,\,\beta_{i}\in[0,1]\}_{i=1}^{M},\\
\{P_{i}\geq0\}_{i=1}^{M+1}
\end{array}}{\mathrm{maximize}} & \sum_{k=1}^{M}R_{k}+\log\left(1+P_{M+1}\bar{\beta}_{M}\right)\label{eq:problem computed cost}\\
\mathrm{s.t.}\,\,\,\,\, & R_{i}\leq\log\left(\frac{1+g_{i}\bar{P}_{i}}{1+g_{i}\bar{P}_{i+1}}\right),\,\, i=1,\ldots,M,\label{eq:problem computed c1}\\
 & C_{i}^{\mathrm{DF}}+\log\left(\frac{1+\bar{P}_{1}\bar{\beta}_{\pi^{-1}(i)}}{1+\bar{P}_{1}\bar{\beta}_{\pi^{-1}(i)-1}}\right)-\log\left(1-\beta_{i}\right)\leq C_{i},\,\, i=1,\ldots,M,\label{eq:problem computed c2}\\
 & \sum_{j=k}^{M}R_{j}\leq\sum_{j=k}^{M}C_{j}^{\mathrm{DF}}+\log\left(\frac{1+\bar{P}_{k}\bar{\beta}_{M}}{1+P_{M+1}\bar{\beta}_{M}}\right),\,\, k=1,\ldots,M,\label{eq:problem computed c3}\\
 & \bar{P}_{1}\leq P,\label{eq:problem computed c4}
\end{align}
\end{subequations}where we have defined variables $\beta_{i}=1/(1+\sigma_{i}^{2})\in[0,1]$
for $i=1,\ldots,M$, the cumulative powers $\bar{P}_{k}=\sum_{j=k}^{M+1}P_{j}$
for $k=1,\ldots,M+1$, the cumulative variables $\bar{\beta}_{i}=\sum_{j=1}^{i}g_{\pi(j)}\beta_{\pi(j)}$
for $i=1,\ldots,M$ and the function $\pi^{-1}(j)$ returns the position
of the index $j\in\{1,\ldots,M\}$ in the ordering $\pi$. The problem
(\ref{eq:problem computed}) is not easy to solve due to the non-convexity
of the constraints (\ref{eq:problem computed c1})-(\ref{eq:problem computed c3}).
In Sec. \ref{subsec:proposed algorithm}, we propose an iterative
algorithm to find a stationary point of the problem (\ref{eq:problem computed}).

\subsection{Proposed Algorithm\label{subsec:proposed algorithm}}

Here we propose an iterative algorithm for finding a stationary point
of problem (\ref{eq:problem computed}). We first simplify the problem
by proving the following lemma.

\begin{lemma}\label{lem:imposing equalities}

Imposing equalities on the constraints (\ref{eq:problem computed c1})
and (\ref{eq:problem computed c2}) induces no loss of optimality.

\begin{proof}\emph{Suppose that the constraints (\ref{eq:problem computed c1})
or (\ref{eq:problem computed c2}) are not satisfied with equality.
Then, we can decrease the transmission powers $P_{1},\ldots,P_{M+1}$
or increase the backhaul usage until the constraints are tight without
decreasing the achievable rate.}

\end{proof}

\end{lemma}

With Lemma \ref{lem:imposing equalities} and some algebraic manipulations,
the problem (\ref{eq:problem computed}) can be written as\begin{subequations}\label{eq:problem cumulative}
\begin{align}
\underset{\{\bar{P}_{i}\geq0\}_{i=1}^{M+1},\{\bar{\beta}_{i},\gamma_{i}\geq0\}_{i=1}^{M}}{\mathrm{minimize}} & \frac{1}{1+P_{M+1}\bar{\beta}_{M}}\prod_{i=1}^{M}\frac{1+g_{i}\bar{P}_{i+1}}{1+g_{i}\bar{P}_{i}}\label{eq:problem reduced cumulative cost}\\
\mathrm{s.t.}\,\,\,\,\,\, & \frac{1+\bar{P}_{M+1}\bar{\beta}_{M}}{2^{\sum_{i=k}^{M}C_{i}}\left(1+\bar{P}_{k}\bar{\beta}_{M}\right)}\prod_{i=k}^{M}\left\{ \frac{\left(1+g_{i}\bar{P}_{i}\right)\left(1+\bar{P}_{1}\bar{\beta}_{\pi^{-1}(i)}\right)}{\gamma_{i}\left(1+g_{i}\bar{P}_{i+1}\right)\left(1+\bar{P}_{1}\bar{\beta}_{\pi^{-1}(i)-1}\right)}\right\} \leq1,\,\, k=1,\ldots,M,\label{eq:problem cumulative c1}\\
 & \frac{1+\bar{P}_{1}\bar{\beta}_{\pi^{-1}(i)}}{2^{C_{i}}\left(1+\bar{P}_{1}\bar{\beta}_{\pi^{-1}(i)-1}\right)\gamma_{i}}\leq1,\,\, i=1,\ldots,M,\label{eq:problem cumulative c2}\\
 & \frac{\bar{P}_{1}}{P}\leq1,\label{eq:problem cumulative c3}\\
 & \frac{\bar{P}_{i+1}}{\bar{P}_{i}}\leq1,\,\frac{\bar{\beta}_{i-1}}{\bar{\beta}_{i}}\leq1,\,\, i=1,\ldots,M,\label{eq:problem cumulative c4}\\
 & \frac{\bar{\beta}_{i}}{g_{\pi(i)}+\bar{\beta}_{i-1}}\leq1,\,\frac{g_{i}\gamma_{i}+\bar{\beta}_{\pi^{-1}(i)}}{g_{i}\gamma_{i}+\bar{\beta}_{\pi^{-1}(i)-1}}\leq1,\,\, i=1,\ldots,M,\label{eq:problem cumulative c5}
\end{align}
\end{subequations}where we characterized the problem over the cumulative
variables $\{\bar{P}_{i}\}_{i=1}^{M+1}$ and $\{\bar{\beta}_{i}\}_{i=1}^{M}$,
and introduced auxiliary variables $\gamma_{i}=1-(\bar{\beta}_{\pi^{-1}(i)}-\bar{\beta}_{\pi^{-1}(i)-1})/g_{i}$
for $i=1,\ldots,M$.

Problem (\ref{eq:problem cumulative}) is not a standard GP \cite{Boyd}
since the denominators in the left-hand side of (\ref{eq:problem cumulative c1}),
(\ref{eq:problem cumulative c2}) and (\ref{eq:problem cumulative c5})
are not monomials. However, the problem is a class of CGP problems
\cite[Sec. 3.2]{Codreanu}, and thus a stationary point of (\ref{eq:problem cumulative})
can be found by applying the homotopy method \cite[Sec. 3.2]{Codreanu},
which solves a sequence of GPs%
\footnote{A GP can be converted into an equivalent convex problem (see \cite[Sec. 4.5.3]{Boyd}
for more detail).%
} obtained by locally approximating the posynomial denominators as
monomial expressions (see, e.g., \cite[Lemma 3.1]{Codreanu}). The
resulting algorithm is summarized in Table Algorithm 1.

\begin{algorithm}
\caption{Homotopy method for problem (\ref{eq:problem cumulative})}

1. Initialize the variables $\{\bar{P}_{i}^{(1)}\geq0\}_{i=1}^{M+1},\{\bar{\beta}_{i}^{(1)}\geq0\}_{i=1}^{M}$
to an arbitrary feasible point and set $n=1$.

2. Update the variables $\{\bar{P}_{i}^{(n+1)}\geq0\}_{i=1}^{M+1},\{\bar{\beta}_{i}^{(n+1)}\geq0\}_{i=1}^{M}$
as a solution of the following GP problem:

\begin{align}
\underset{\{\bar{P}_{i}^{(n+1)}\geq0\}_{i=1}^{M+1},\{\bar{\beta}_{i}^{(n+1)},\gamma_{i}\geq0\}_{i=1}^{M}}{\mathrm{minimize}} & \frac{1}{f\left(P_{M+1}^{(n+1)}\bar{\beta}_{M}^{(n+1)},P_{M+1}^{(n)}\bar{\beta}_{M}^{(n)}\right)}\prod_{i=1}^{M}\frac{1+g_{i}\bar{P}_{i+1}^{(n+1)}}{f\left(g_{i}\bar{P}_{i}^{(n+1)},g_{i}\bar{P}_{i}^{(n)}\right)}\label{eq:problem reduced cumulative cost-1}\\
\mathrm{s.t.}\,\,\,\,\,\, & \prod_{i=k}^{M}\left\{ \frac{\left(1+g_{i}\bar{P}_{i}^{(n+1)}\right)\left(1+\bar{P}_{1}^{(n+1)}\bar{\beta}_{\pi^{-1}(i)}^{(n+1)}\right)}{\gamma_{i}f\left(g_{i}\bar{P}_{i+1}^{(n+1)},g_{i}\bar{P}_{i+1}^{(n)}\right)f\left(\bar{P}_{1}^{(n+1)}\bar{\beta}_{\pi^{-1}(i)-1}^{(n+1)},\bar{P}_{1}^{(n)}\bar{\beta}_{\pi^{-1}(i)-1}^{(n)}\right)}\right\} \nonumber \\
 & \times\frac{1+\bar{P}_{M+1}^{(n+1)}\bar{\beta}_{M}^{(n+1)}}{2^{\sum_{i=k}^{M}C_{i}}f\left(\bar{P}_{k}^{(n+1)}\bar{\beta}_{M}^{(n+1)},\bar{P}_{k}^{(n)}\bar{\beta}_{M}^{(n)}\right)}\leq1,\,\, k=1,\ldots,M,\nonumber \\
 & \frac{1+\bar{P}_{1}^{(n+1)}\bar{\beta}_{\pi^{-1}(i)}^{(n+1)}}{2^{C_{i}}\gamma_{i}f\left(\bar{P}_{1}^{(n+1)}\bar{\beta}_{\pi^{-1}(i)-1}^{(n+1)},\bar{P}_{1}^{(n)}\bar{\beta}_{\pi^{-1}(i)-1}^{(n)}\right)}\leq1,\,\, i=1,\ldots,M,\nonumber \\
 & \frac{\bar{P}_{1}^{(n+1)}}{P}\leq1,\nonumber \\
 & \frac{\bar{P}_{i+1}^{(n+1)}}{\bar{P}_{i}^{(n+1)}}\leq1,\,\frac{\bar{\beta}_{i-1}^{(n+1)}}{\bar{\beta}_{i}^{(n+1)}}\leq1,\,\, i=1,\ldots,M,\nonumber \\
 & \frac{\bar{\beta}_{i}^{(n+1)}}{g_{\pi(i)}f\left(\bar{\beta}_{i-1}^{(n+1)}/g_{\pi(i)},\bar{\beta}_{i-1}^{(n)}/g_{\pi(i)}\right)}\leq1,\,\, i=1,\ldots,M,\nonumber \\
 & \frac{g_{i}\gamma_{i}+\bar{\beta}_{\pi^{-1}(i)}^{(n+1)}}{g_{i}f\left(\bar{\beta}_{\pi^{-1}(i)-1}^{(n+1)}/g_{i},\bar{\beta}_{\pi^{-1}(i)-1}^{(n)}/g_{i}\right)}\leq1,\,\, i=1,\ldots,M,\nonumber
\end{align}
where the function $f(s,\hat{s})$ is a monomial function of $s$
defined as \cite[Lemma 3.1]{Codreanu}
\begin{equation}
f(s,\hat{s})=c(\hat{s})s^{a(\hat{s})}\label{eq:monomial LB-1}
\end{equation}
with $a(\hat{s})=\hat{s}(1+\hat{s})^{-1}$ and $c(\hat{s})=\hat{s}^{-a}(1+\hat{s})$.

3. Stop if some convergence criterion is satisfied. Otherwise, set
$n\leftarrow n+1$ and go to Step 2.
\end{algorithm}

\section{Special Cases \label{sec:Special-Cases}}

Here we discuss some relevant special cases of the proposed scheme.

\subsection{Compress-and-Forward}

If we impose that the encoder uses only the highest layer $X_{M+1}$,
i.e., $X=\sqrt{P}X_{M+1}$ in lieu of the more general (\ref{eq:broadcast coding}),
the proposed hybrid scheme reduces to a pure CF scheme with successive
decoding as studied in \cite{Park}\cite{WeiYu}. Optimization of the
test channels $\beta_{1},\ldots,\beta_{M}$ under this assumption
and given ordering $\pi$ can be simplified to
\begin{align}
\underset{\beta_{1},\ldots,\beta_{M}\geq0}{\mathrm{maximize}} & \log\left(1+P\sum_{j=1}^{M}g_{j}\beta_{j}\right)\label{eq:problem CF only}\\
\mathrm{s.t.}\,\,\,\, & \log\left(\frac{1+P\bar{\beta}_{i}}{1+P\bar{\beta}_{i-1}}\right)-\log\left(1-\beta_{\pi(i)}\right)\leq C_{\pi(i)},\,\, i=1,\ldots,M,\nonumber
\end{align}
whose solutions $\beta_{1}^{\mathrm{opt}},\ldots,\beta_{M}^{\mathrm{opt}}$
are directly given, using Lemma \ref{lem:imposing equalities}, as
\begin{equation}
\beta_{\pi(i)}^{\mathrm{opt}}=\frac{\left(2^{C_{\pi(i)}}-1\right)\left(1+P\bar{\beta}_{i-1}\right)}{2^{C_{i}}\left(1+P\bar{\beta}_{i-1}\right)+Pg_{\pi(i)}},\,\, i=1,\ldots,M.
\end{equation}

\subsection{Decode-and-Forward}

The DF strategy is a special case of the proposed hybrid relaying
scheme obtained by fixing $\beta_{1}=\ldots=\beta_{M}=0$ and $P_{M+1}=0$.
A similar approach was studied in \cite[Sec. V-B]{Xue} for $M=2$
assuming Gaussian channels for relay-to-destination links. A stationary
point of the problem can be obtained by adopting the homotopy method
in Algorithm 1 with minor modifications. As an interesting special
case, we consider DF with single-layer transmission in which multi-layer
transmission is not leveraged.

Using single-layer transmission, the following rate is achievable
by optimizing the selection of the transmitted layer:
\begin{equation}
\max_{i\in\mathcal{M}}\min\left\{ \log\left(1+g_{i}P\right),\sum_{j=i}^{M}C_{j}\right\} .\label{eq:problem DF SL}
\end{equation}
We remark that in (\ref{eq:problem DF SL}) we have used the fact,
as in the more general result of Lemma \ref{lem:achievability}, that
all relays $i,\ldots,M$ are able to decode message $W_{i}$ and thus
the message can be distributed across the backhaul links in order
to be delivered to the destination.

\section{Numerical Results\label{sec:Numerical-Results}}

\begin{figure}
\centering\includegraphics[width=10.5cm,height=8.1cm]{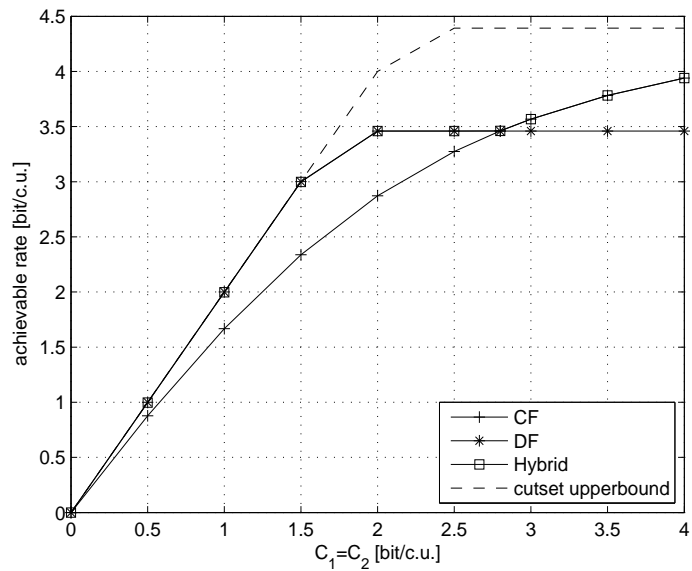}

\caption{\label{fig:graph as C sym}Achievable rates versus the backhaul capacity
$C_{1}=C_{2}$ in a symmetric network with $M=2$, $P=0\,\mathrm{dB}$
and $g_{1}=g_{2}=10\,\mathrm{dB}$.}
\end{figure}

In this section, we present numerical results to investigate the advantage
of the proposed multi-layer transmission scheme with hybrid relaying
studied in Sec. \ref{sec:Proposed}-\ref{sec:optimization} as compared
to the more conventional schemes reviewed in Sec. \ref{sec:Special-Cases}.
For reference, we also compare the achievable rates with the cutset
upper bound \cite[Theorem 1]{Cover}
\begin{equation}
R_{\mathrm{cutset}}=\min_{\mathcal{S}\subseteq\{1,\ldots,M\}}\left\{ \sum_{j\in\mathcal{S}}C_{j}+\log\left(1+P\sum_{j\in\mathcal{S}^{c}}g_{j}\right)\right\} .\label{eq:cutset bound}
\end{equation}
For ease of interpretation, we focus on the case with two relays,
i.e., $M=2$. We mark single-layer schemes with the label 'SL' and
multi-layer schemes with 'ML'. For CF related schemes, the optimal
ordering $\pi^{\mathrm{opt}}$ in problem (\ref{eq:problem original})
was found via exhaustive search and was observed to be $\pi=(1,2)$
for all the simulated cases.

In Fig. \ref{fig:graph as C sym}, we examine the performance in a
symmetric setting by plotting the rate versus the backhaul capacities
$C_{1}=C_{2}$ when $P=0\,\mathrm{dB}$ and $g_{1}=g_{2}=10\,\mathrm{dB}$.
It is seen that in this symmetric set-up, the optimized hybrid scheme
ends up reducing to either the DF or the CF strategy at small and
large backhaul capacity, respectively. Note that we have not distinguished
between the single-layer and multi-layer strategies in the figure
since they showed the same performance when the relays experience
the same fading power, i.e., $g_{1}=g_{2}$. This is expected since
multi-layer strategies are relevant only when the two relays have
different decoding capabilities.

\begin{figure}
\centering\includegraphics[width=10.5cm,height=8.1cm]{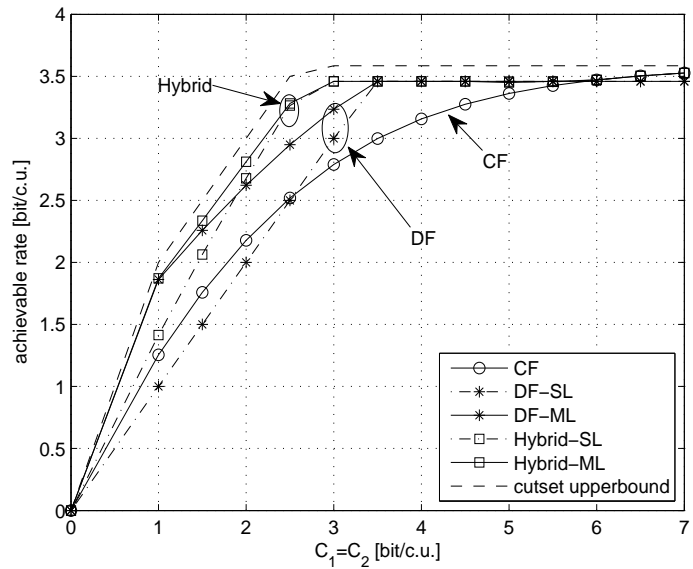}

\caption{\label{fig:graph as C asym}Achievable rates versus the backhaul capacity
$C_{1}=C_{2}$ per relay with $M=2$, $P=0\,\mathrm{dB}$ and $[g_{1},g_{2}]=[0,10]\,\mathrm{dB}$.}
\end{figure}

In Fig. \ref{fig:graph as C asym}, we observe the performance versus
the backhaul capacity $C_{1}=C_{2}$ with $P=0\,\mathrm{dB}$ and
asymmetric channel powers $[g_{1},g_{2}]=[0,10]\,\mathrm{dB}$. Unlike
the symmetric setting in Fig. \ref{fig:graph as C sym}, the multi-layer
strategy is beneficial compared to the single-layer (SL) transmission
for both DF and Hybrid schemes%
\footnote{Not being based on relay decoding, CF operates only with one layer.%
}. Moreover, unlike the setting of Fig. \ref{fig:graph as C sym},
the hybrid relaying strategy shows a performance advantage with respect
to all other schemes. This is specifically the case for intermediate
values of the backhaul capacities $C_{1}=C_{2}$. It should also be
mentioned that, as $C_{1}=C_{2}$ increases, the performance of DF
schemes is limited by the capacity of the better decoder, namely $\log_{2}(1+10)=3.46$
bit/c.u., while CF, and thus also the hybrid strategy, are able, for
$C_{1}=C_{2}$ large enough, to achieve the cutset bound.

\begin{figure}
\centering\includegraphics[width=10.5cm,height=8.1cm]{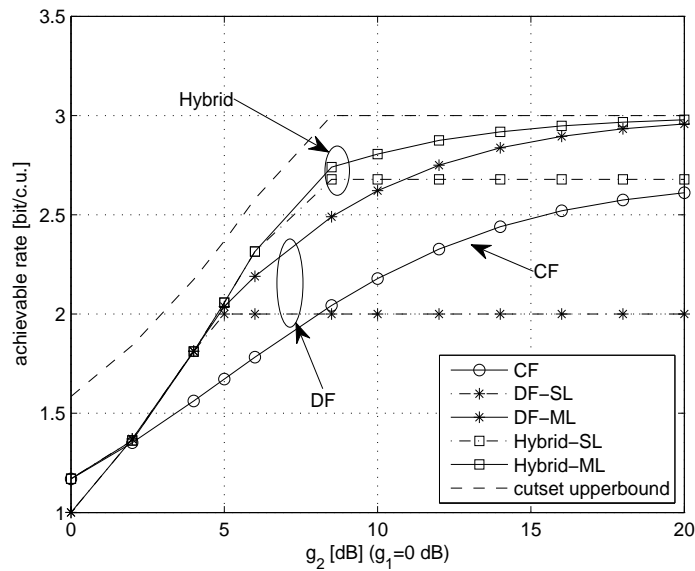}

\caption{\label{fig:graph as G2 asym}Achievable rates versus the channel power
$g_{2}$ with $M=2$, $P=0\,\mathrm{dB}$, $g_{1}=0\,\mathrm{dB}$
and $C_{1}=C_{2}=2$ bit/c.u..}
\end{figure}

Finally, in Fig. \ref{fig:graph as G2 asym}, we plot the achievable
rates versus the channel power $g_{2}$ of the better relay when $P=0\,\mathrm{dB}$,
$g_{1}=0\,\mathrm{dB}$ and $C_{1}=C_{2}=2$ bit/c.u.. As expected,
the performance gain of multi-layer transmission over the single-layer
schemes is more pronounced as $g_{2}$ increases, since a better channel
to relay 2 allows to support larger rates for both rates of both DF
layers. In fact, single-layer transmission uses only the DF layer
decoded exclusively by relay 2 according to (\ref{eq:problem DF SL}).
For the same reason, the rate of single-layer DF is limited by the
backhaul capacity $C_{2}$ of relay 2. Moreover, hybrid relaying is
advantageous over all conventional schemes for intermediate values
of $g_{2}$.

\section{Conclusions\label{sec:Conclusions}}

We have studied transmission and relaying techniques for the relay
channels with multiple out-of-band relays, which are connected to
the destination via orthogonal finite-capacity backhaul links. We
proposed a novel transmission and relaying strategies whereby multi-layer
transmission is used at the encoder and hybrid DF-CF relaying is adopted
at the relays. The multi-layer transmission is designed so as to adaptively
leverage the different decoding capabilities of the relays and to
enable the hybrid relaying strategy. As a result, the proposed multi-layer
strategy is different from the classical broadcast coding approach
of \cite{Shamai BC}, which aims at coping with uncertain fading conditions
at the transmitter (see also \cite{Khandani} for an application to
a multi-relay setting).

We aimed at maximizing the achievable rate, which is formulated as
a non-convex problem. However, based on the observation that the problem
falls in the class of so called Complementary Geometric Programs (CGPs),
we have proposed an iterative algorithm based on the homotopy method
which attains a stationary point of the problem. From numerical results,
it was shown that the proposed multi-layer transmission with the hybrid
relaying strategy outperforms more conventional decode-and-forward,
compress-and-forward and single-layer strategies, especially in the
regime of moderate backhaul capacities and asymmetric channel gains
from the source to the relays.

\appendices

\section{Proof of Lemma \ref{lem:achievability}}\label{appendix:proof achievability}

Here, we show that conditions (\ref{eq:backhaul DF}) are sufficient
for correct decoding of messages $W_{1},\ldots,W_{M}$ at the decoder.
To see this, we observe that the destination, when decoding messages
$W_{1},\ldots,W_{M}$, can be regarded as the decoder of a multiple
access channel with $M$ sources. Specifically, source $k$ has messages
$W_{1},\ldots,W_{k}$ for $k=1,\ldots,M$ and has two inputs to the
channel to the destination, namely the signal $X_{k}$ and the information
sent at rate $C_{k}^{\mathrm{DF}}$ on the noiseless backhaul link.
We denote the latter as $T_{k}$, where $T_{k}\in\{1,\ldots,2^{C_{k}^{\mathrm{DF}}}\}$
so that the overall channel input of the source $k$ is given by $\tilde{X}_{k}=(X_{k},T_{k})$.
The destination observes $\hat{Y}_{\mathcal{M}}$ and $T_{1},\ldots,T_{M}$.
We emphasize that both $X_{k}$ and $T_{k}$ in $\tilde{X}_{k}$ depend
on all messages $W_{1},\ldots,W_{k}$.

As a result, we have an equivalent multiple access channel in which
each source has a specific subset of all the messages and a hierarchy
exists among the sources so that source $k$ has all the messages
also available to sources $1,\ldots,k-1$. Therefore, using the results
in \cite{Bruyn}\cite{Gunduz}, the following conditions guarantee
correct decoding of messages $W_{1},\ldots,W_{M}$
\begin{equation}
\sum_{j=k}^{M}R_{j}\leq I\left(\tilde{X}_{\{k,\ldots,M\}};\hat{Y}_{\mathcal{M}},T_{\{1,\ldots,M\}}|\tilde{X}_{\{1,\ldots,k-1\}}\right),\label{eq:inequality appendix 1}
\end{equation}
for $k=1,\ldots,M$. The achievability of rates (\ref{eq:inequality appendix 1})
is ensured for any joint distribution of the inputs $\{\tilde{X}_{k}\}_{k=1}^{M}$
\cite{Bruyn}\cite{Gunduz}. To proceed, we take $\tilde{X}_{k}$ to
be independent according to the discussion around (\ref{eq:broadcast coding}),
and also take $X_{k}$ to be independent of $T_{k}$ for all $k=1,\dots,M$.
It is not hard to see that this choice maximizes the mutual informations
in (\ref{eq:inequality appendix 1}). Under these assumptions, we
can write the right-hand side of (\ref{eq:inequality appendix 1})
as
\begin{align}
 & I\left(X_{\{k,\ldots,M\}},T_{\{k,\ldots,M\}};\hat{Y}_{\mathcal{M}},T_{\{1,\ldots,M\}}|X_{\{1,\ldots,k-1\}},T_{\{1,\ldots,k-1\}}\right)\\
= & I\left(X_{\{k,\ldots,M\}};\hat{Y}_{\mathcal{M}}|X_{\{1,\ldots,k-1\}}\right)+H\left(T_{\{k,\ldots,M\}}\right)\nonumber \\
= & I\left(X_{\{k,\ldots,M\}};\hat{Y}_{\mathcal{M}}|X_{\{1,\ldots,k-1\}}\right)+\sum_{j=k}^{M}C_{j}^{\mathrm{DF}},\nonumber
\end{align}
by the chain rule for mutual informations \cite[Theorem 2.5.2]{Cover}.
This proves that inequalities (\ref{eq:inequality appendix 1}) reduce
to (\ref{eq:backhaul DF}) with the given choices.

\end{document}